\documentclass
[10pt,showpacs,letterpaper,runinaddress,prd]{revtex4}%
\usepackage{amsfonts}
\usepackage{amsmath}
\usepackage{amssymb}
\usepackage{graphicx}%
\begin{document}
\title{Strangelet propagation and cosmic ray flux}
\author{Jes Madsen}
\affiliation{Department of Physics and Astronomy, University of Aarhus, DK-8000 \AA rhus C, Denmark}

\begin{abstract}
The galactic propagation of cosmic ray strangelets is described and the
resulting flux is calculated for
a wide range of parameters as a prerequisite for strangelet searches in lunar
soil and with an Earth orbiting magnetic spectrometer, AMS-02. 
While the inherent uncertainties are large, flux predictions 
at a measurable level are obtained for
reasonable choices of parameters if strange quark matter is absolutely stable.
This allows a direct test of the strange matter hypothesis.

\end{abstract}
\date{November 17, 2004}
\maketitle

\section{Introduction}

At densities slightly above nuclear matter density quark matter composed of
up, down, and strange quarks in roughly equal numbers (called strange quark
matter) may be absolutely stable, i.e. stronger bound than iron 
\cite{Bodmer:1971we,Chin:1979yb,Witten:1984rs,Farhi:1984qu}. In spite of
two decades of scrutiny, neither theoretical calculations, nor experiments or
astrophysical observations have been able to settle this issue
(for reviews, see \cite{Madsen:1998uh,Weber:2004kj}). If strange
quark matter is absolutely stable it would have important consequences for
models of ``neutron stars'', which would then
most likely all be quark stars (strange stars
\cite{Witten:1984rs,Haensel:1986qb,Alcock:1986hz,Glendenning:1997wn,Weber:1999qn}). 
It would also give rise to a
significant component of strangelets (lumps of strange quark matter) in cosmic
rays, and in fact the search for cosmic ray strangelets may be the most direct
way of testing the stable strange matter hypothesis. A significant flux of
cosmic ray strangelets could exist due to strange matter release from binary
collisions of strange stars \cite{Madsen:1989pg,Friedman:1990qz}. 
In the following the propagation of such cosmic
ray strangelets in the Galaxy is discussed, and the resulting differential and
total flux is calculated as a function of various parameters within a
simple, but physically transparent propagation model. The results are
of interest for the coming cosmic ray experiment on The International Space
Station, Alpha Magnetic Spectrometer (AMS-02)
\cite{Aguilar:2002ad,AMS:2004,Sandweiss:2004bu,Madsen:2001bw}, 
as well as for a lunar soil strangelet search.

\section{Strangelet properties}

Strangelets are lumps of strange quark matter with baryon number $A$ in the
possible range from a few to a few times $10^{57}$ (the term strangelet is
sometimes used only for lumps with baryon number below $10^{6}-10^{7}$; here
the term will be used for lumps of all $A$). The upper limit is the
gravitational instability limit for truly macroscopic bulk quark matter in the
form of strange stars. Given absolute stability of bulk strange quark matter,
a lower limit on $A$ for stable strangelets follows from the fact that surface
tension and curvature energy destabilize small quark lumps relative to bulk
matter. However, the stability of bulk strange quark matter as well as the
value of the minimal $A$ depends on poorly determined model parameters such as
the strange quark mass, the bag constant and the strong coupling constant, so
in the following stability is simply assumed, and $A$ is treated as a free parameter.

Several models of strange quark matter have been discussed in the literature,
and it has been realized that different phases of strange quark matter may be
energetically favorable depending on the values of relevant parameters. Common
for all of the corresponding types of strangelets are masses per baryon below
the mass of nuclei (by assumption of stability), and a very low charge-to-mass
ratio compared with nuclei due to the near cancellation of up, down and
strange quark charges when these quarks appear in nearly equal numbers.

In the following strangelet masses are assumed to be near the upper limit for
stability,%
\begin{equation}
mc^{2}=m_{0}c^{2}A\approx0.93\text{GeV}A.
\end{equation}
Slightly lower masses are possible, and the assumption made here neglects the
fact that the strangelet mass per baryon decreases slightly with $A$ in model
calculations. These effects are of negligible consequence relative to the far
larger uncertainties in other parameter choices below.

Strangelet charges, $Z$, are significantly lower than the $Z\approx A/2$ known
from nuclei, and this is the most important experimental signature for
strangelet detection. For the sake of example, most results in the figures
below assume strangelets to be color-flavor locked with a charge 
\cite{Madsen:2001fu}
\begin{equation}
Z=0.3A^{2/3}.
\end{equation}
This assumes a strange quark mass of 150 MeV (the charge of color-flavor
locked strangelets is proportional to the strange quark mass 
\cite{Madsen:2001fu}). Color-flavor
locked quark matter is electrically neutral in bulk
\cite{Rajagopal:2000ff}, because the pairing of
quarks is optimized when all three quark flavors have equal Fermi momenta and
thereby equal bulk number densities. However, a depletion of massive strange
quarks at the surface of a finite quark matter lump leads to a net positive
charge proportional to the surface area or $A^{2/3}$
\cite{Madsen:2000kb,Madsen:2001fu}. Strangelets without
pairing (\textquotedblleft ordinary\textquotedblright\ strangelets) have
charge \cite{Heiselberg:1993dc}
\begin{align}
Z  &  =0.1A\\
Z  &  =8A^{1/3}%
\end{align}
for $A\ll 700$ and $A\gg 700$ respectively, again assuming a strange quark
mass of 150 MeV (here the charge is proportional to $m_{s}^{2}$
\cite{Heiselberg:1993dc,Farhi:1984qu,Berger:1986ps}).

Strangelets may be neutralized by electrons and form unusual ions or atoms. At
very high charges ($Z\gg10^{2}$), the strangelets are automatically partly
neutralized by electrons from the excitation of the vacuum
\cite{Madsen:2002iw}, so the net charge
increases slower with mass than indicated in the relations above. For most of
the results below the masses are below the range where such effects become
important. Also, for most of the calculations the cosmic ray strangelets have
energies where full ionization is a good approximation, but the effective
charge is smaller than $Z$ at small velocities,
c.f.~Eq.~(\ref{eq:effch}) below.

\section{Strangelet production}

The production of strangelets is speculative, but could happen when two
strange stars in a binary system spiral towards each other due to loss of
orbital energy in the form of gravitational radiation. If strange quark matter
is absolutely stable (the assumption in this paper for stable cosmic ray
strangelets to exist) all compact stars are likely to be strange stars
\cite{Madsen:1989pg,Friedman:1990qz},  and
therefore the galactic coalescence rate will be the one for double neutron
star binaries recently updated in \cite{Kalogera:2003tn} based on available
observations of binary pulsars to be $83.0_{-66.1}^{+209.1}$ Myr$^{-1}$ at a
95\% confidence interval, thus of order one collision in our Galaxy every
3,000--60,000 years.

Each of these events involve a phase of tidal disruption of the stars as they
approach each other before the final collision. During this stage small
fractions of the total mass may be released from the binary system in the form
of strange quark matter. No realistic simulation of such a collision involving
two strange stars has been performed to date. Newtonian and
semirelativistic simulations of the
inspiral of strange stars and black holes do exist 
\cite{Lee:2002nk,Kluzniak:2002dm,Prakash:2003em}, but the physics is too
different from the strange star-strange star collision to be of guidance.
Simulations of binary neutron star collisions, depending on orbital and other
parameters, lead to the release of anywhere from $10^{-5}-10^{-2}M_{\odot}$,
where $M_{\odot}$ denotes the solar mass, corresponding to a total mass
release in the Galaxy of anywhere from $10^{-10}-3\times 10^{-6}M_{\odot}$
per year
with the collision rate above. Given the high stiffness of the equation of
state for strange quark matter, strange star-strange star collisions should
probably be expected to lie in the low end of the mass release range, so the
canonical input for the following calculations is a galactic production rate
of
\begin{equation}
\overset{\cdot}{M}=10^{-10}M_{\odot}\text{yr}^{-1}.
\end{equation}

All strangelets released are assumed to have a single baryon number, $A$. This
is clearly a huge oversimplification, but there is no way of calculating the
actual mass spectrum to be expected. As demonstrated in \cite{Madsen:2001bw} the
quark matter lumps originally released by tidal forces are macroscopic in size
(when estimated from a balance between quark matter surface tension and tidal
forces), but subsequent collisions will lead to fragmentation, and under the
assumption that the collision energy is mainly used to compensate for the
extra surface energy involved in making smaller strangelets, it was argued
that a significant fraction of the mass released from binary strange star
collisions might ultimately be in the form of strangelets with $A\approx
10^{2}-10^{4}$, though these values are strongly parameter dependent.

Fortunately, most of the total flux results derived for cosmic ray strangelets
below are such that values for some given $A$ are valid as a lower limit for
the flux for a fixed total strangelet mass injection if strangelets are
actually distributed with baryon numbers below $A$.

\section{Strangelets in cosmic rays: acceleration and propagation}

Apart from an unusually high $A/Z$-ratio compared to nuclei, strangelets would
in many ways behave like ordinary cosmic ray nuclei. For example, the most
likely acceleration mechanism would be Fermi acceleration in supernova shocks
resulting in a rigidity spectrum at the source which is a powerlaw in rigidity
as described below. Due to the high strangelet rigidity, $R$, at fixed
velocity ($R\equiv pc/Ze= Am_{0}c^{2}\gamma (\beta ) \beta/Ze$, where $p$
is the strangelet momentum, $\beta\equiv v/c$,
and $\gamma=(1-\beta^{2})^{-1/2}$) 
strangelets are more efficiently injected into an accelerating shock
than are nuclei with $A/Z\approx2$ (c.f. discussion of nuclei in
\cite{Gieseler}), and one may expect that most strangelets passed by a
supernova shock will take part in Fermi acceleration.

The time scales for strangelet acceleration, energy loss, spallation and escape
from the Galaxy are all short compared to the age of the Milky Way Galaxy.
This makes it reasonable to assume that cosmic ray strangelets are described
by a steady state distribution, i.e. as a solution to a propagation equation
of the form
\begin{equation}
\frac{dN}{dt}=0
\end{equation}
where $N(E,x,t)dE$ is the number density of strangelets at position $x$ and
time $t$ with energy in the range $[E,E+dE]$. In standard form 
(e.g.\ \cite{Longair:1994wu,Longair:1992ze}; given the significant
uncertainty in input parameters in the present investigation, a simple but
physically transparent model for
strangelet propagation is chosen) $\frac{dN}{dt}$
is given by the following sum of a source term from supernova acceleration, a
diffusion term, loss terms due to escape from the Galaxy, energy loss, decay,
spallation, and a term to describe reacceleration of strangelets due to
passage of new supernova shock waves,
\begin{equation}
\frac{dN}{dt}=\frac{\partial N}{\partial t}|_{\text{source}}+D\nabla
^{2}N+\frac{\partial N}{\partial t}|_{\text{escape}}+\frac{\partial}{\partial
E}[b(E)N]+\frac{\partial N}{\partial t}|_{\text{decay}}+\frac{\partial
N}{\partial t}|_{\text{spallation}}+\frac{\partial N}{\partial t}%
|_{\text{reacceleration}}.
\end{equation}
The individual terms will be defined and discussed in the following
subsections.

\subsection{Source term}

The strangelet spectrum after acceleration in supernova shocks is assumed to
be a standard powerlaw in rigidity as derived from observations of ordinary
cosmic rays,%
\begin{equation}
g(R)=\frac{\alpha-1}{R_{\min}^{1-\alpha}-R_{\max}^{1-\alpha}}R^{-\alpha},
\end{equation}
where rigidity $R$ is measured in GV, powerlaw index $\alpha\approx2.2$, and
the normalization is such that $\int_{R_{\min}}^{R_{\max}}g(R)dR=1$. The
minimal rigidity is assumed to be given by the speed of a typical supernova
shock wave, $\beta_{SN}\approx0.005,$ where $v\equiv\beta c$ is the speed, so
$R_{\min}=\gamma(\beta_{SN})\beta_{SN}Am_{0}c^{2}/Ze\approx5$MV$A/Z$. The
maximal rigidity from acceleration in supernova shocks, $R_{\max},$ is of
order $10^{6}$ GV, but the actual number is irrelevant since $R_{\max}\gg
R_{\min}$ and $g(R)$ decreases rapidly with increasing $R$.

The source term is normalized according to a total production rate of
$\overset{\cdot}{M}=10^{-10}M_{\odot}$yr$^{-1}$ of baryon number $A$
strangelets spread evenly in an effective galactic volume (see the following
subsection) of $V=1000$ kpc$^{3}$, giving a total source term%

\begin{equation}
G(R)=\frac{\overset{\cdot}{M}}{VAm_{0}}g(R)
\end{equation}
or in terms of energy (using $dE/dR=Ze\beta$ and $G(R)dR=G(E)dE$)%
\begin{equation}
\frac{\partial N}{\partial t}|_{\text{source}}\equiv G(E)=\frac{G(R(E))}%
{Ze\beta}.
\end{equation}

\subsection{Spatial diffusion and escape}

The terms $D\nabla^{2}N+\frac{\partial N}{\partial t}|_{\text{escape}}$, where
$D$ is the diffusion coefficient, describe cosmic ray diffusion in real space
and eventual escape from the confining magnetic field of the Galaxy. Charged
cosmic rays are spiralling along field lines in the weak galactic magnetic
field, but due to the very irregular structure of the field, the particles
scatter on magnetic \textquotedblleft impurities\textquotedblright, and the
motion is best described in terms of diffusion. From studies of nuclei it is
known, that cosmic rays are confined to move in a region significantly larger
than the galactic disk, where most of the stars and interstellar matter are
located. A typical value for the effective galactic volume confining cosmic
rays is $V=1000$ kpc$^{3}$ as chosen above. We will make a standard leaky box
approximation, assuming $D=0$ and $\frac{\partial N}{\partial t}%
|_{\text{escape}}=-N/\tau_{\text{escape}}$, where $\tau_{\text{escape}%
}(A,Z,E)$ is the average escape time from an otherwise homogeneous
distribution in the galactic volume, $V$. From studies of cosmic ray nuclei
the average column density of material, $\xi_{\text{escape}}$, passed by
nuclei before escape from the Galaxy is known as a function of rigidity, $R$,
as
\begin{equation}
\xi_{\text{escape}}=\xi_{\text{0}}\left(  \frac{R}{R_{0}}\right)  ^{\delta},
\end{equation}
where $\xi_{\text{0}}=12.8$ g/cm$^{2}$, $R_{0}=4.7$GV, $\delta=0.8$ for
$R<R_{0}$, and $\delta=-0.6$ for $R>R_{0}$. Strangelets are sufficiently
similar to nuclei for the same distribution of column density to be assumed.
The corresponding time scale is given by $\tau=\xi/(\rho v)$, where $\rho$ is
the density of the interstellar medium (largely atomic and molecular
hydrogen). With $n$ denoting the average hydrogen number density per cubic
centimeter ($n\approx0.5$ when averaging over denser regions in the galactic
plane and dilute regions in the magnetic halo), the resulting escape time
scale for strangelets is%
\begin{equation}
\tau_{\text{escape}}=\frac{8.09\times10^{6}\text{y}}{n\beta}\left(  \frac
{R}{R_{0}}\right)  ^{\delta}.
\end{equation}

\subsection{Energy loss}

The term in the propagation equation $\frac{\partial}{\partial E}[b(E)N]$,
describes the influence of energy loss processes. The energy loss rate
$b(E)\equiv-dE/dt$ can be treated as a sum of ionization losses (from
interaction with neutral hydrogen atoms and molecules), Coulomb losses (from
interaction with ionized hydrogen), and pion production losses from inelastic
collisions at high relativistic $\gamma$-factor (threshold at $\gamma=1.3$).
Since Coulomb losses are generally negligible compared to the other two
mechanisms, only ionization and pion production will be considered.

\subsubsection{Pion production loss}

In accordance with the relations for nuclei given in \cite{Schlickeiser} we
assume%
\begin{equation}
\frac{dE}{dt}|_{\text{pion}}=-1.82\times10^{-7}nA^{0.53}0.72\gamma
^{1.28}H(\gamma-1.3)\text{ eV/s},
\end{equation}
where $H$ is the Heaviside step function taking account of the pion production
threshold, and the Lorentz factor is $\gamma=(1-\beta^{2})^{-1/2}.$

\subsubsection{Ionization loss}

The energy loss due to ionization is given by \cite{Schlickeiser}
\begin{equation}
\frac{dE}{dt}|_{\text{ionization}}=-1.82\times10^{-7}nY(Z,\beta)^{2}%
[1+0.0185\ln(\beta)H(\beta-0.01)]\frac{2\beta^{2}}{10^{-6}+2\beta^{3}}\text{
eV/s},
\end{equation}
where the effective charge $Y$ is smaller than $Z$ at small velocities as
described by%
\begin{equation}
Y(Z,\beta)=Z[1-1.034\exp(-137\beta Z^{-0.688})].
\label{eq:effch}
\end{equation}
This relation is based on fits to intermediate and
high-$Z$ nuclei, and it is not expected to work at very 
low values of $\beta$. However it is
sufficient for the present purpose since the strangelet cosmic ray spectrum
at the lowest energies is dramatically reduced by solar modulation anyway.
At speeds close to the speed of light the ionization loss is simply
proportional to $nZ^{2}$.

\subsection{Decay}

The existence of a measurable cosmic ray strangelet flux requires strangelets
to be stable. Thus, it is assumed that $\frac{\partial N}{\partial
t}|_{\text{decay}}=0.$

\subsection{Spallation}

Like nuclei strangelets have a roughly geometrical cross section (proportional
to $A^{2/3}$) for spallation in collisions with interstellar matter
(hydrogen). The corresponding spallation time scale is taken to be%
\begin{equation}
\tau_{\text{spallation}}=\frac{2\times10^{7}\text{y}}{n\beta}A^{-2/3},
\end{equation}
where the normalization comes from data for nuclei at kinetic energy above 2
GeV. At low kinetic energy the cross section for nuclei (and presumably
strangelets) can vary somewhat due to resonances etc, but such complications
will be neglected here, since the detailed physics is unknown in the case of
strangelets. We have also neglected the slight reduction in geometrical area
of strangelets relative to nuclei due to their slightly larger density. The
largest uncertainty in the treatment of spallation is the fact that
strangelets (like nuclei) are not always completely destroyed in a spallation
reaction. In addition to nucleons and nuclei smaller strangelets may result
from this type of reaction, but we are ignorant of the physics to an extent
where it is impossible to include this effective feed-down to lower $A$ in a
meaningful manner. Therefore, spallation is assumed to be a process destroying
strangelets, i.e.%
\begin{equation}
\frac{\partial N}{\partial t}|_{\text{spallation}}=-\frac{N}{\tau
_{\text{spallation}}}.
\end{equation}
This leads to an overall underestimate of the strangelet flux.

\subsection{Reacceleration}

Strangelet energies are redistributed according to the propagation equation.
Some leave the Galaxy or are destroyed by spallation. Occasionally strangelets
get a new kick from a passing supernova shock, and in a first approximation
they regain the source term relative distribution of rigidity, $g(R)$. The
time scale between supernova shock waves passing a given position in
interstellar space is of order $\tau_{SN}\approx10^{7}$ y. This scale is
comparable to or larger than the time scales for energy loss, spallation, and
escape from the Galaxy, so reacceleration of cosmic ray strangelets has only a
moderate influence on the steady state distribution. By adding energy (on
average) to the particles it actually reduces the total flux of strangelets
somewhat because higher energies make destruction and escape more likely. The
reacceleration mechanism has been taken into account by adding terms to the
right hand side of the propagation equation of the form%
\begin{equation}
\frac{\partial N}{\partial t}|_{\text{reacceleration}}=-\frac{N}{\tau_{SN}%
}+\frac{\int N(E)dE}{\tau_{SN}}g(R)\frac{dR}{dE},
\end{equation}
where the first term describes how particles are taken out of the spectrum on
a time scale $\tau_{SN}$, and the second term reintroduces these particles
with the source term rigidity power law distribution.

The importance of reacceleration relative to the source term from new
strangelets is parametrized by $K$, which is given as
\begin{equation}
K=\left[  \frac{\int N(E)dE}{\tau_{SN}}\right]  /\left[  \frac{\overset{\cdot
}{M}}{VAm_{0}}\right]  \text{.}%
\end{equation}
Due to the large value of $\tau_{SN}$ (of order $10^{7}$ years, which is
comparable to or bigger than the other relevant time scales), $K$ is typically
rather small, and therefore reacceleration plays a minor role (factor of 2 or
less for typical parameters).

\subsection{The energy distribution}

Introducing the terms discussed above the steady state equation $\frac{dN}%
{dt}=0$ leads to the following differential equation for $N(E)$%
\begin{equation}
b(E)\frac{dN}{dE}=\frac{N(E)}{\tau(E,A,Z)}-\left[  1+K\right]  G(E),
\end{equation}
where%
\begin{equation}
1/\tau(E,A,Z)=1/\tau_{\text{escape}}+1/\tau_{\text{spallation}}+1/\tau
_{SN}+1/\tau_{\text{loss}},
\end{equation}
with $1/\tau_{\text{loss}}\equiv -db(E)/dE.$

\subsection{The interstellar flux}

Given a solution for $N(E)$ the corresponding flux in the \textquotedblleft
average\textquotedblright\ interstellar medium with energies from $[E,E+dE]$
is given by%
\begin{equation}
F_{E}(E)dE=\frac{\beta c}{4\pi}N(E)dE\text{,}%
\end{equation}
and the corresponding flux in terms of rigidity is%
\begin{equation}
F_{R}(R)dR=Ze\beta F_{E}(E(R))dR
\end{equation}
(using $dE/dR=Ze\beta$).

\subsection{Solar modulation}

Like other charged cosmic ray particles strangelets are influenced by the
solar wind when entering the inner parts of the Solar System. The detailed
interactions are complicated, but as demonstrated for nuclei in
\cite{Gleeson:1968}, a good fit to the solar modulation of
the cosmic ray spectrum can be given in terms of a potential model, where the
charged particle climbs an electrostatic potential of order $\Phi=500$ MeV
(the value changes by a factor of less than 2 during the 11 (22) year solar
cycle). This effectively reduces the cosmic ray energy by $|Z|\Phi$ relative
to the value in interstellar space, and at the same time the flux is reduced
by the relative reduction in particle momentum squared, so that the modulated
spectrum is%
\begin{equation}
F_{\text{mod}}(E)=\left(  \frac{R(E)}{R(E+|Z|\Phi)}\right)  ^{2}%
F_{E}(E+|Z|\Phi)\text{.}%
\end{equation}
Solar modulation significantly suppresses the flux of charged
cosmic rays at energies below a few GeV. The solar modulation effectively
works like a smooth cut-off in flux below kinetic energy of order $Z\Phi$.
Since strangelets have a high mass-to-charge ratio they are nonrelativistic
at these energies, which correspond to rigidities in GV of $R_{\text{GV}%
}\approx(A/Z)^{1/2}\Phi_{500}^{1/2}$, where $\Phi_{500}=\Phi/(500$MeV$).$

\subsection{Geomagnetic cutoff}

For cosmic rays to reach the Earth or an Earth-orbiting detector like the
Alpha Magnetic Spectrometer on the International Space Station, the energies
have to exceed the geomagnetic cutoff rigidity, which is a function of
detector position, and for an orbiting observatory like AMS the value varies
from 1--15 GV as a function of time. Notice that the geomagnetic cutoff
rigidity for low mass strangelets is comparable to or higher than the solar
modulation cutoff, whereas high mass strangelets experience solar modulation
already at rigidities above the geomagnetic cutoff.

For a non-magnetic body like the Moon, there is no corresponding cutoff, and
the total flux is given by $F_{\text{mod}}$.

\section{Results}

\subsection{Special cases}

While the general solution of the propagation equation requires numerical
integration, several limits can be treated analytically and provide a physical
understanding of the full numerical solutions that follow below. The special
cases (disregarding solar modulation and geomagnetic cutoff) can be divided
according to the relative importance of the different time scales,
$\tau_{\text{escape}}$, $\tau_{\text{spallation}}$, $\tau_{SN}$ and
$\tau_{\text{loss}}$ (or rather $|\tau_{\text{loss}}|$, since the energy loss
time scale as defined above is negative at low and high energies, describing a
net increase in number of particles).

When one of the time scales is significantly smaller than the others at a given
energy (rigidity), the corresponding process dominates the physics. The
relative importance of the processes depends on strangelet properties $A,Z$,
on the density of interstellar hydrogen $n$ (though most processes have the
same $n$-dependence), and of course on the strangelet energy, $E$ (or
equivalently rigidity, $R$, speed, $\beta$, or Lorentz-factor, $\gamma$). For
most strangelet masses and charges, $\tau_{SN}$ is larger than one or
more of the other time scales. Typically, energy loss dominates at low energy,
spallation at intermediate $E$, and escape from the Galaxy at the highest
energies. Figure 1 compares the different time scales for color-flavor
locked strangelets with $Z=8, A\approx 138$. The
energy loss time scale plotted is $|\tau_{\text{loss}}|$, whereas
$\tau_{\text{loss}}$ itself is negative to the right and left of the
parabola-like section of $|\tau_{\text{loss}}|$ between rigidities of order
$0.5$ to $10$ GV. The total time scale is negative below $0.5$ GV, since it is
almost equal to the energy loss time scale in this regime.

\subsubsection{Low energy; energy loss domination}

At low energies $\tau\approx\tau_{\text{loss}}$, and the propagation equation
reduces to%
\begin{equation}
\frac{d}{dE}\left[  b(E)N(E)\right]  \approx-G(E),
\end{equation}
which can be integrated to give%
\begin{equation}
N(E)\approx-\frac{\int_{\infty}^{E}G(E^{\prime})dE^{\prime}}{b(E)}.
\end{equation}

At the very lowest energies for which the source term $g(R)=0$ (this happens
when $R<R_{\min}$), the integral in the numerator is a (negative) constant,
and $N(E)\propto1/b(E)$. Above $R_{\min}$, $N(E)\propto(R/R_{\min}%
)^{-1.2}/b(E)$. The corresponding flux in terms of rigidity includes an
additional $\beta^{2}$-dependence proportional to $R^{2}$ in the
nonrelativistic domain. In this domain energy loss by ionization dominates. If
the strangelet charge were constant, the flux would be constant at very low
$R$, followed by regimes with decreasing flux proportional to $R^{-1.2}$,
increasing flux proportional to $R^{1.8}$, and another regime with the flux
decreasing like $R^{-1.2}$ when approaching the realm of relativistic
strangelets (however, except for very low $(A,Z)$, the energy loss domination
is only relevant for nonrelativistic strangelets).

Figure 2 shows the differential flux as a function of rigidity for
$Z=8, A\approx 138$ (as in Figure 1). 
The low energy domain of the full numerical
solution (marked ``Interstellar flux'') shows
a decrease rather than constant flux at the far left end, followed by a
slightly steeper decrease for $R>R_{\min}$, an increase, and again a decrease
in flux. In general, the flux is higher at low $R$ and the decrease with $R$
steeper (increase with $R$ less steep) than would be expected from the
arguments above. This is because the effective charge $Y(Z,\beta)$ is less
than $Z$ for nonrelativistic strangelets and approaches $0$ for low $R$.
(In fact, the effective charge as defined here becomes negative at
finite $R$. This is of course unphysical, but reflects the fact that the
energy loss due to ionization becomes negligible).

\subsubsection{Intermediate energy; spallation domination}

At intermediate and high energies energy loss processes can be neglected, and
thus $b(E)\approx0$, corresponding to $|\tau_{\text{loss}}|\approx\infty$. For
most parameter choices this shift takes place at rigidities somewhat below the
solar modulation cutoff discussed above. At intermediate energies the spectrum
is determined by the strangelet spallation time, $\tau\approx\tau
_{\text{spallation}}$, and the propagation equation is approximately given by%
\begin{equation}
N(E)\approx\left[  1+K\right]  G(E)\tau_{\text{spallation}}(n,\beta,A)
\end{equation}
with $\tau_{\text{spallation}}=2\times10^{7}$y$n^{-1}\beta^{-1}A^{-2/3}.$

Assuming also $K=0$ (equivalent to $\tau_{\text{SN}}\gg\tau_{\text{spallation}%
}$) gives the approximate result%
\begin{equation}
F_{R}(R)=2.34\times10^{5}\text{m}^{-2}\text{yr}^{-1}\text{sterad}%
^{-1}\text{GV}^{-1}A^{-0.467}Z^{-1.2}R_{\text{GV}}^{-2.2}\Lambda,
\end{equation}
or for the total flux above rigidity $R$
\begin{equation}
F(>R)=1.95\times10^{5}\text{m}^{-2}\text{yr}^{-1}\text{sterad}^{-1}%
A^{-0.467}Z^{-1.2}R_{\text{GV}}^{-1.2}\Lambda.
\end{equation}

In both cases the results are proportional to
\begin{equation}
\Lambda=\left(  \frac{\beta_{\text{SN}}}{0.005}\right)  ^{1.2}\left(
\frac{0.5\text{cm}^{-3}}{n}\right)  \left(  \frac{\overset{\cdot}{M}}%
{10^{-10}M_{\odot}\text{yr}^{-1}}\right)  \left(  \frac{1000\text{kpc}^{3}}%
{V}\right)  \left(  \frac{930\text{MeV}}{m_{0}c^{2}}\right)  .
\label{lambda}
\end{equation}
Notice that the differential strangelet spectrum keeps the source term slope,
$G(R)\propto R^{-2.2}$. This is also clearly seen in Figure 2, where the
spallation regime corresponds to rigidities from a few GV to a few tens of GV.
Figure 3 shows the corresponding curves for the integrated flux ($F(>R)$).

\subsubsection{High energy; escape time domination}

The intermediate energy domain is replaced by the high energy domain when
$\tau_{\text{escape}}\leq\tau_{\text{spallation}}$. Except for very low $A$
this happens when $R>R_{0}\left[  2.5A^{-2/3}\right]  ^{-1/0.6}\approx
1.0$GV$A^{1.11}$, or $E>1.0$GeV$ZA^{1.11}\beta^{-1}$. At high energies the
spectrum is determined by the confinement time of strangelets in the Galaxy,
$\tau\approx\tau_{\text{escape}}$, and the propagation equation is
approximately given by%
\begin{equation}
N(E)\approx G(E)\tau_{\text{escape}}(n,\beta,R).
\end{equation}
For semirelativistic or relativistic strangelets with $\beta\approx1$,
$\tau_{\text{escape}}\propto R^{-0.6}$, so the spectrum is steepened from the
source term $R^{-2.2}$ to $R^{-2.8}$.

Again assuming $K=0$ (equivalent to $\tau_{\text{SN}}\gg\tau_{\text{escape}}$)
gives the approximate result%
\begin{equation}
F_{R}(R)=2.40\times10^{5}\text{m}^{-2}\text{yr}^{-1}\text{sterad}%
^{-1}\text{GV}^{-1}A^{0.2}Z^{-1.2}R_{\text{GV}}^{-2.8}\Lambda,
\end{equation}
and for the total flux above rigidity $R$
\begin{equation}
F(>R)=1.33\times10^{5}\text{m}^{-2}\text{yr}^{-1}\text{sterad}^{-1}%
A^{0.2}Z^{-1.2}R_{\text{GV}}^{-1.8}\Lambda.
\end{equation}
Comparison with Figures 2 and 3 shows good agreement at high $R$ (the small
difference is due to reacceleration).

\subsubsection{Rule of thumb for the total flux}

Since the astrophysical input parameters in the present calculations are
uncertain at the order of magnitude level, relations for the total flux of
strangelets hitting the Earth or Moon accurate within a factor of $2$ (for
fixed input parameters) are useful. As indicated above solar modulation
effectively cuts off the strangelet flux at rigidities of order $R_{\text{SM}%
}\approx(A/Z)^{1/2}\Phi_{500}^{1/2}$GV, which is in the part of the spectrum
where the strangelet flux is governed by spallation. The total flux hitting
the Moon or Earth is therefore roughly given by%
\begin{equation}
F_{\text{total}}\approx2\times10^{5}\text{m}^{-2}\text{yr}^{-1}\text{sterad}%
^{-1}A^{-0.467}Z^{-1.2}\max[R_{\text{SM}},R_{\text{GC}}]^{-1.2}\Lambda,
\end{equation}
depending on whether solar modulation or geomagnetic cutoff dominates. In the
case of solar modulation domination (always relevant for the Moon, and
relevant for AMS as long as $R_{\text{SM}}>R_{\text{GC}}$) one obtains%
\begin{equation}
F_{\text{total}}\approx2\times10^{5}\text{m}^{-2}\text{yr}^{-1}\text{sterad}
^{-1}A^{-1.067}Z^{-0.6}\Phi_{500}^{-0.6}\Lambda.
\label{appflux}
\end{equation}
For strangelets obeying the CFL mass-charge relation $Z=0.3A^{2/3}$ this
becomes
\begin{align}
F_{\text{total}}\approx & 4\times10^{5}\text{m}^{-2}\text{yr}^{-1}\text{sterad}
^{-1}A^{-1.467}\Phi_{500}^{-0.6}\Lambda\\ \approx & 2.8\times10^{4}\text{m}^{-2}
\text{yr}^{-1}\text{sterad}^{-1}Z^{-2.2}\Phi_{500}^{-0.6}\Lambda,
\end{align}
which reproduces the numerical results to within 20\% for $Z>10$ and to within
a factor of a few even for small $Z$, where the assumptions of nonrelativistic
strangelets and spallation domination both are at the limit of being valid.

\subsection{Numerical results}

Figures 2 and 3 show the interstellar strangelet fluxes as a function of
rigidity, as well as the fluxes reduced by solar modulation. Differential
fluxes as well as integrated fluxes are displayed for strangelets with charge
$Z=8$, and baryon number $A\approx 138$ given by the charge-mass
relation for color-flavor locked strangelets.

Figures 4 and 5 show the total strangelet flux as a function of strangelet
charge and baryon number respectively, assuming the CFL-relation for the
charge. Figures 6 and 7 show similar results, assuming the charge-mass
relation of ``ordinary'' strangelets. 
Figures 4--7 do not include geomagnetic cutoff, so they are
relevant for the Moon, whereas fluxes could be reduced for a detector in Earth
orbit, depending primarily on geomagnetic latitude. 
Notice the nice agreement between the numerical calculations and the
approximation to $F_{\text{total}}$ given above, except at very low
charge.

The strangelet flux calculated for $\Lambda$ of order unity
is high enough to be of interest for
various upcoming experimental searches, and at the same time small
enough to agree with previous searches which have given
upper limits or shown marginal evidence for signatures consistent with
strangelets (see \cite{Sandweiss:2004bu} for an overview).
As stressed several times above, many parameters are
uncertain at the order of magnitude level. The scaling with these
parameters is indicated where relevant. In particular this is true for
the overall normalization of the strangelet flux as expressed via the
parameter $\Lambda$ (Eq.\ (\ref{lambda})), 
whereas the relative behavior of the differential
flux is less uncertain. 

Apart from the uncertainty in parameters within the picture discussed
here, one cannot rule out the possibility that some of the basic
assumptions need to be modified. In addition to strangelet production
in strange star collisions, it has been suggested that a (possibly
small) flux of strangelets may be a direct outcome of the Type II
supernova explosions \cite{Benvenuto:1989hw}, where strange stars form. A
detailed quantitative treatment of this mechanism for cosmic ray strangelet
production does not exist.
The treatment in the present paper does not include such additional
strangelet production mechanisms, and therefore the flux predictions are
conservative. Another assumption that leads to a conservative lower
limit on the flux is that spallation is assumed to destroy strangelets
completely. At least for low-energy collisions one would expect that
fragments of strangelets would survive, but the input physics for
performing a realistic strangelet spallation study is not sufficiently
well known, and therefore the conservative assumption of complete
destruction was made in the present study. A numerical simulation of
strangelet propagation in Ref.~\cite{Medina-Tanco:1996je} assumed
stripping of nucleons rather than complete destruction of strangelets in
interstellar collisions and studied two specific examples for the
mass and energy of strangelets injected into the interstellar medium. 
Several other assumptions made in that numerical study differ from those of the
current investigation, and therefore a direct comparison of the results
is not possible, except for the reassuring fact that both studies are
consistent with the possibility of a significant, measurable strangelet flux  
in our part of the Galaxy.

But ultimately the question of whether strangelets 
exist in cosmic rays is an experimental issue.

\section{Conclusion}

Strangelet fluxes have been calculated numerically and compared to approximate
values derived in various parameter regimes. The total strangelet flux
reaching the Moon or a detector in Earth orbit is in a regime that could be
within experimental reach, and therefore provide a crucial test of the
hypothesis of absolutely stable strange quark matter.

\acknowledgments
This work was supported by the Danish Natural Science Research Council.
I appreciate the hospitality of DOE's Institute for Nuclear Theory in
Seattle, where part of this work was carried out.
I thank Alexei Chikanian, Evan Finch, Richard Majka, and Jack Sandweiss
for comments and suggestions, and 
Jonas M\o ller Larsen for joint efforts in an early phase of this work.

\begin{figure}
\begin{center}
\includegraphics[
keepaspectratio,
width=4.5in,angle=270
]%
{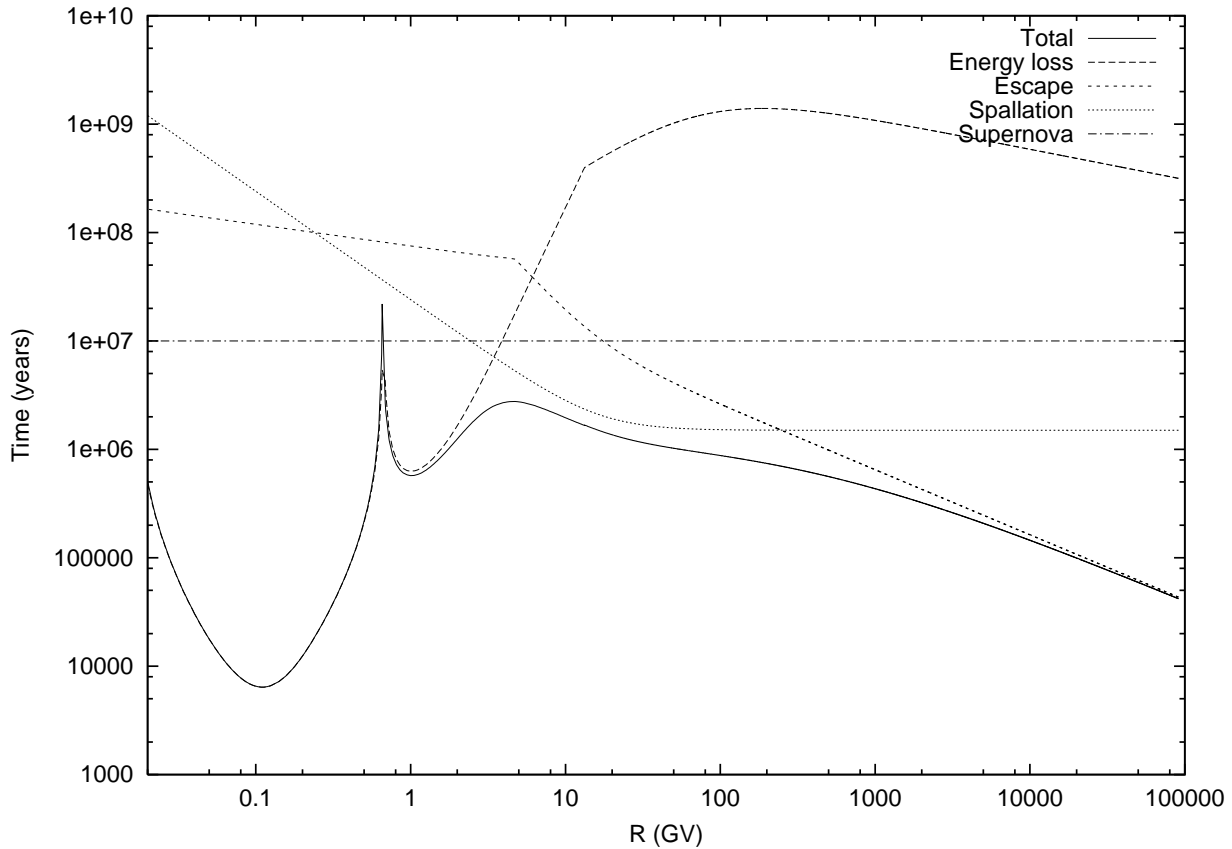}
\caption{Characteristic time scales (in years) as a function of rigidity
(in GV) for the processes dominating the propagation of strangelets:
Energy loss ($|\tau_{\protect\text {loss}}|$; long-dashed),
escape from the Galaxy ($\tau_{\text {escape}}$; short-dashed), 
spallation ($\tau_{\text {spallation}}$; dotted), and
reacceleration in supernova shocks ($\tau_{SN}$; dotted-dashed). The
total time scale, $\tau$, is shown by the solid curve. All time scales are
shown positive, but $\tau_{\text {loss}}$ is actually negative
outside the parabola-like regime from 0.5--10 GV (it becomes positive
again at very low values of $R$). The total time scale is dominated by
energy loss below approximately 3~GV and follows the sign change of
$\tau_{\text {loss}}$ there. For intermediate rigidities spallation is
most important, and at high $R$ escape from the Galaxy is the fastest
and therefore dominating process. The example given assumes color-flavor
locked strangelets with $Z=8$ ($A\approx 138$)
and neglects solar modulation and geomagnetic cutoff.}
\label{figure 1}
\end{center}
\end{figure}

\begin{figure}
\begin{center}
\includegraphics[
keepaspectratio,
width=4.5in,angle=270
]%
{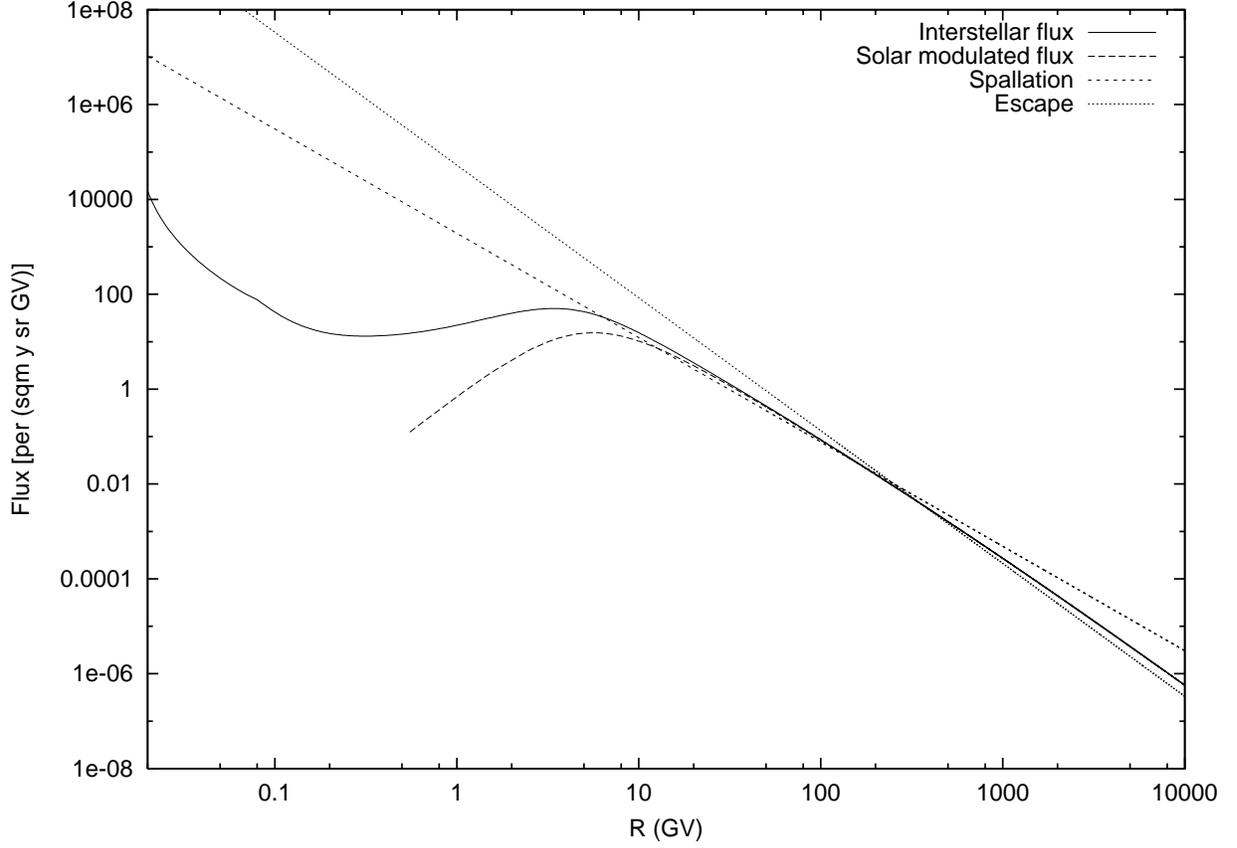}
\caption{The differential strangelet flux as a function of rigidity 
assuming color-flavor
locked strangelets with $Z=8, A\approx 138$.
The solid curve shows the ``interstellar flux'', i.e.\ the average flux
in the Galaxy without inclusion of solar modulation and geomagnetic
cutoff. The long-dashed curve includes solar modulation. Geomagnetic cutoff
may eliminate the flux below rigidities of a few GV for a detector in
Earth orbit, whereas no such effect exists for strangelets hitting the
Moon or other non-magnetic objects. Short-dashed and dotted curves
show the effects of spallation and galactic escape domination
respectively. It is seen that these processes explain the behavior in
specific rigidity regimes, as discussed in the text.}
\label{figure 2}
\end{center}
\end{figure}

\begin{figure}
\begin{center}
\includegraphics[
keepaspectratio,
width=4.5in,angle=270
]%
{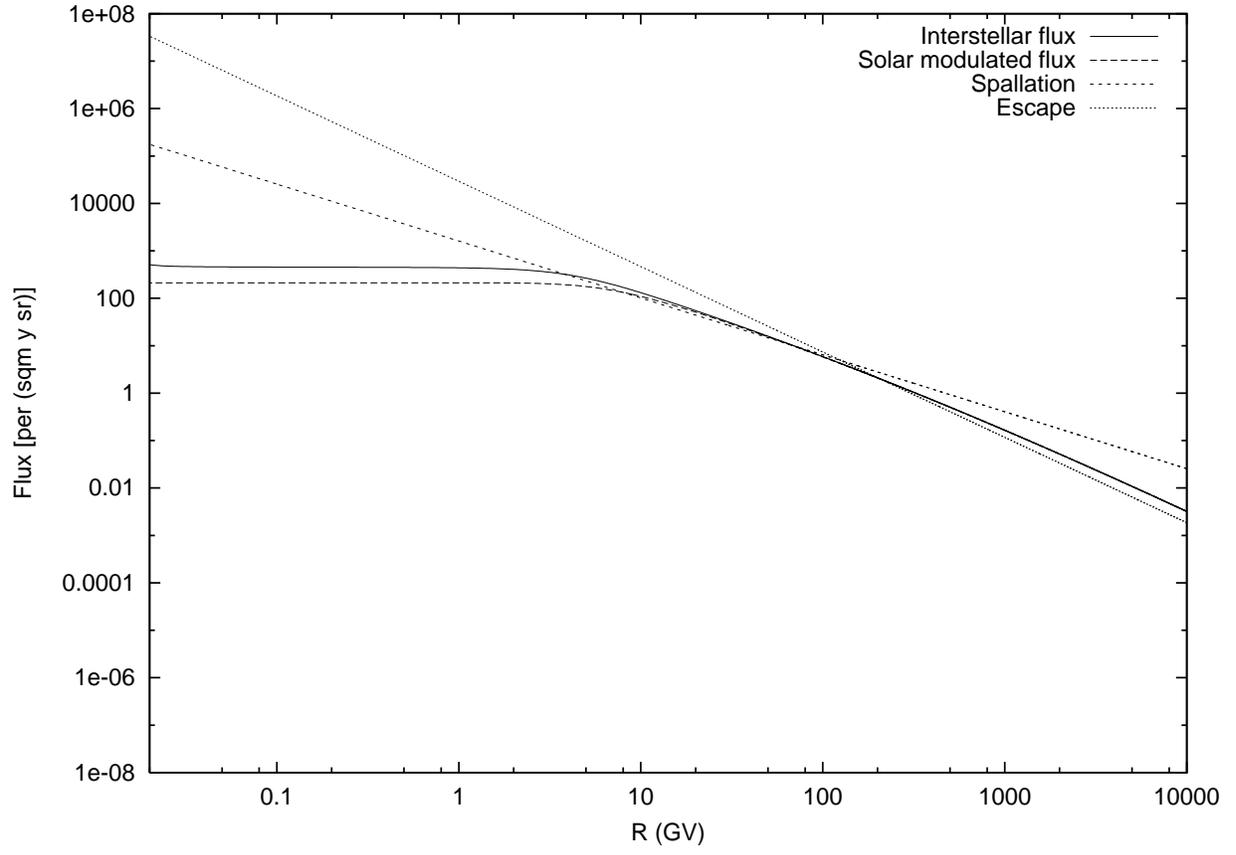}
\caption{The integrated strangelet flux
at rigidity higher than $R$. Parameters and curves as in
Figure 2.}
\label{figure 3}
\end{center}
\end{figure}

\begin{figure}
\begin{center}
\includegraphics[
keepaspectratio,
width=4.5in,angle=270
]%
{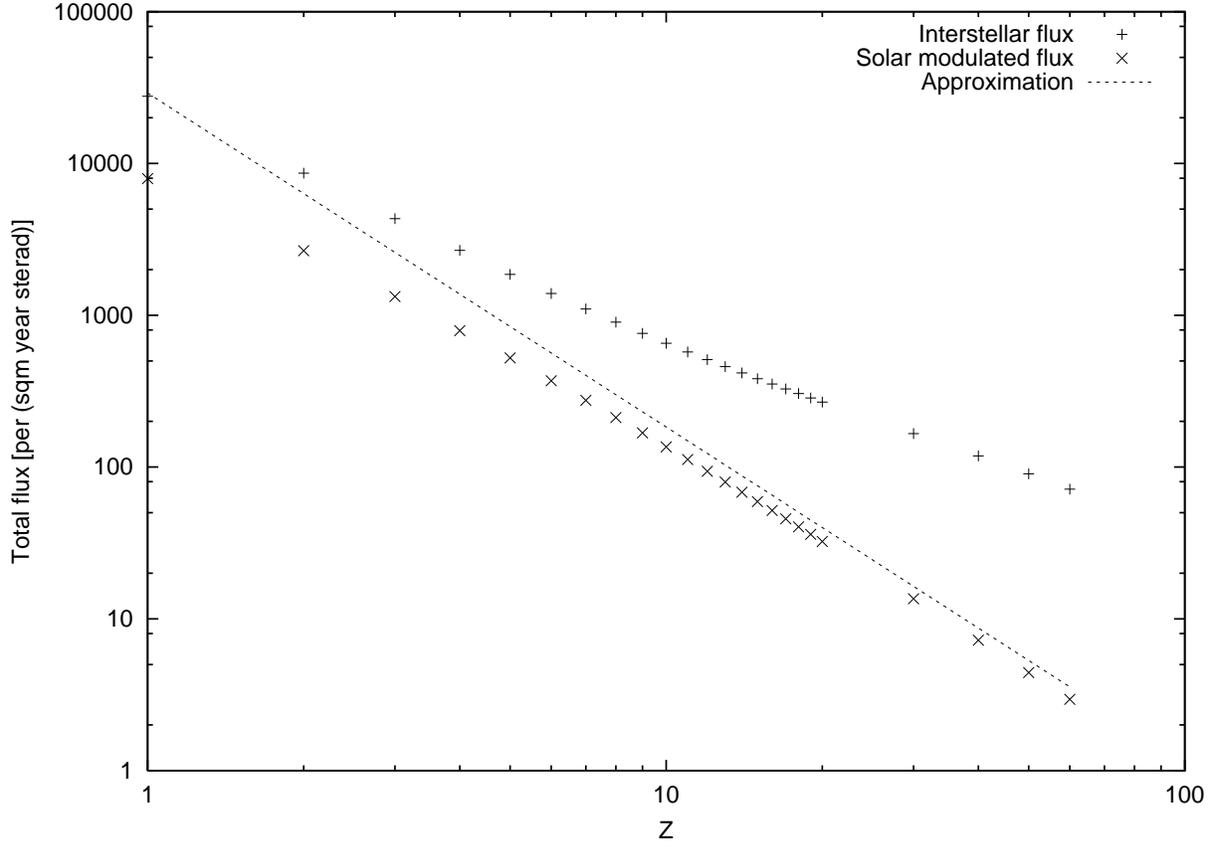}
\caption{The total strangelet flux as a function of charge for
interstellar conditions and near Earth taking account of solar
modulation, but not geomagnetic cutoff. The approximation in 
Eq.\ (\protect\ref{appflux}) is
seen to reproduce the local strangelet flux very well, except at very
low $Z$. Strangelets are assumed to follow the CFL charge-mass
relation.}
\label{figure 4}
\end{center}
\end{figure}

\begin{figure}
\begin{center}
\includegraphics[
keepaspectratio,
width=4.5in,angle=270
]%
{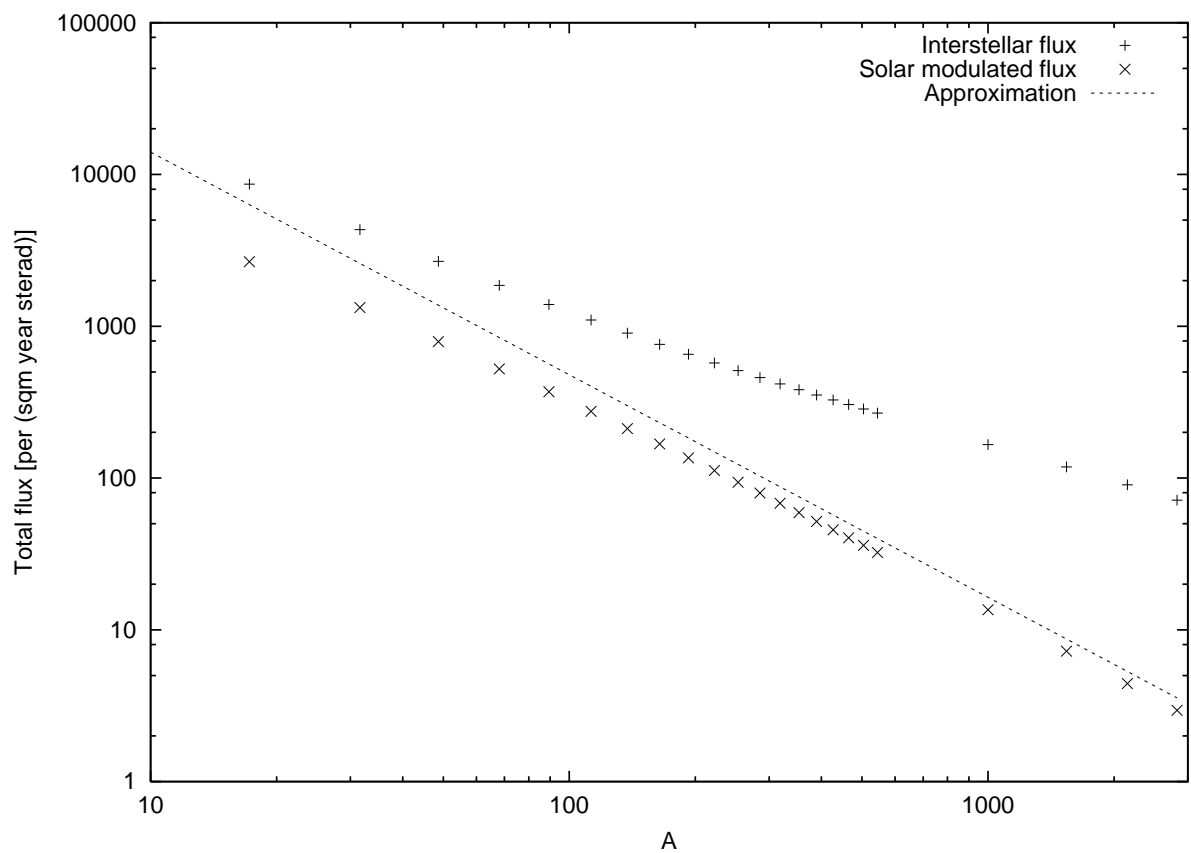}
\caption{As Figure 4, except that the total flux is now shown as
a function of baryon number.}
\label{figure 5}
\end{center}
\end{figure}

\begin{figure}
\begin{center}
\includegraphics[
keepaspectratio,
width=4.5in,angle=270
]%
{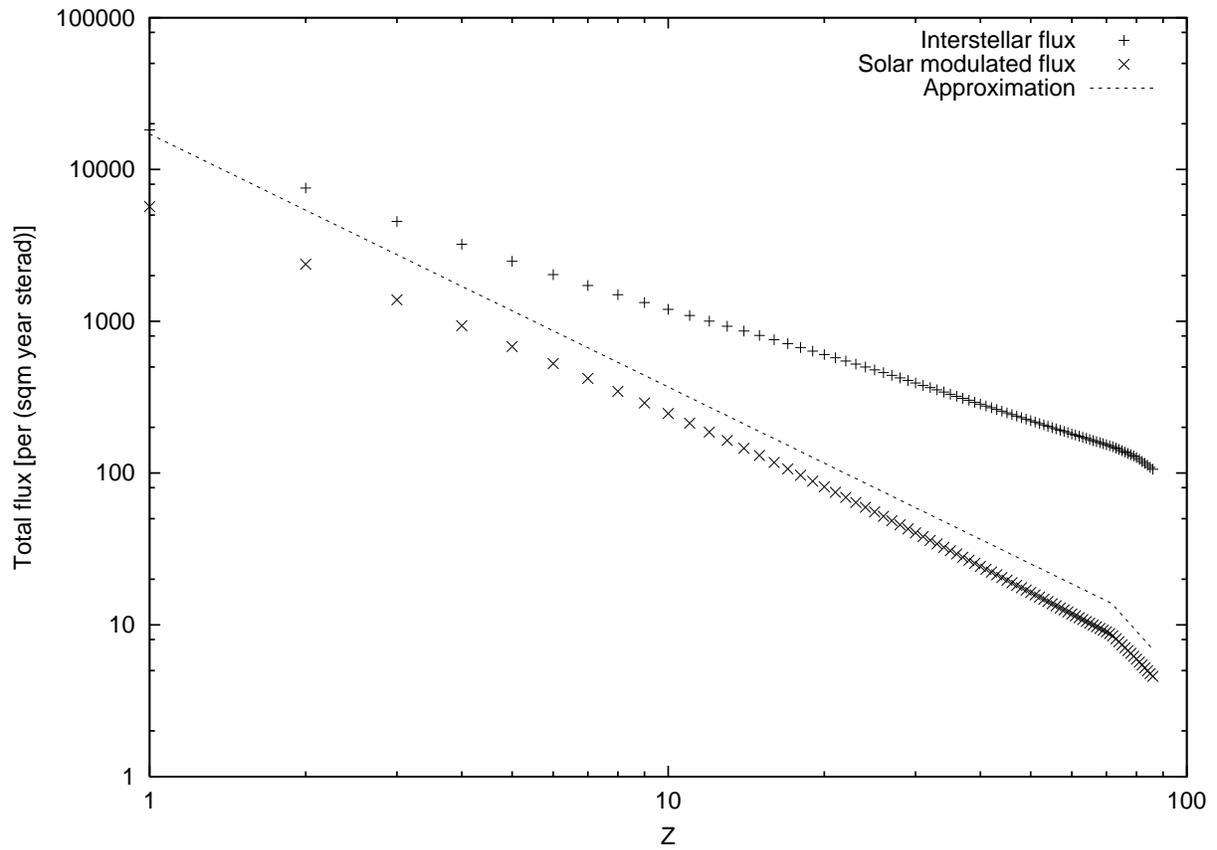}
\caption{As Figure 4, except that the charge-mass relation for ordinary
strangelets rather than color-flavor locked strangelets is assumed. The
change in slope near $Z=70$ is due to the change in slope of the $Z$-$A$
relation for ordinary strangelets.}
\label{figure 6}
\end{center}
\end{figure}

\begin{figure}
\begin{center}
\includegraphics[
keepaspectratio,
width=4.5in,angle=270
]%
{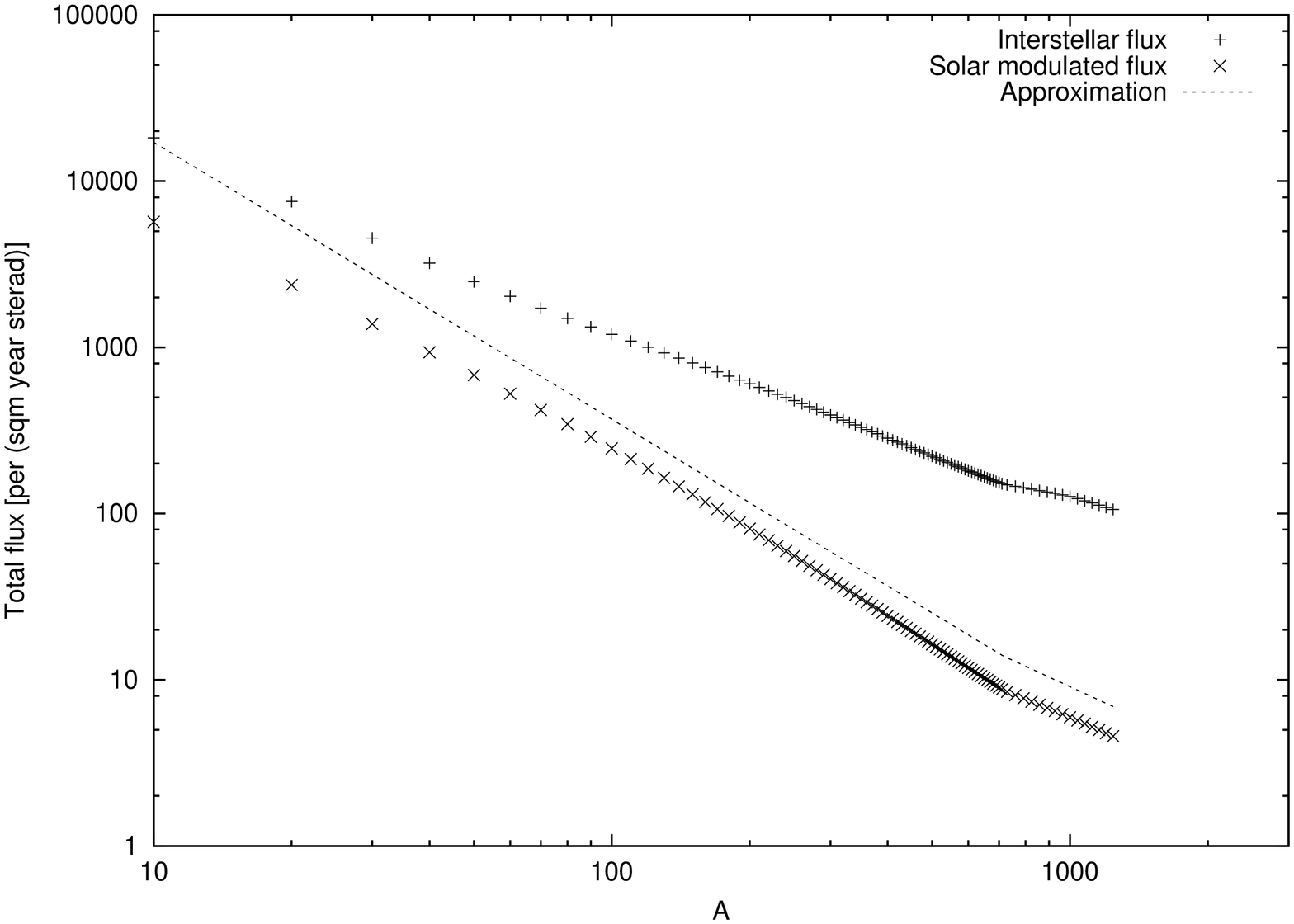}%
\caption{As Figure 6, except that the flux of ordinary strangelets 
is shown as a function of baryon number.}
\label{figure 7}
\end{center}
\end{figure}


\begin{thebibliography}{9} 

%\cite{Bodmer:1971we}
\bibitem{Bodmer:1971we}
A.~R.~Bodmer,
%``Collapsed Nuclei,''
Phys.\ Rev.\ D {\bf 4}, 1601 (1971).
%%CITATION = PHRVA,D4,1601;%%

%\cite{Chin:1979yb}
\bibitem{Chin:1979yb}
S.~A.~Chin and A.~K.~Kerman,
%``Possible Longlived Hyperstrange Multi - Quark Droplets,''
Phys.\ Rev.\ Lett.\  {\bf 43}, 1292 (1979).
%%CITATION = PRLTA,43,1292;%%

%\cite{Witten:1984rs}
\bibitem{Witten:1984rs}
E.~Witten,
%``Cosmic Separation Of Phases,''
Phys.\ Rev.\ D {\bf 30}, 272 (1984).
%%CITATION = PHRVA,D30,272;%%

%\cite{Farhi:1984qu}
\bibitem{Farhi:1984qu}
E.~Farhi and R.~L.~Jaffe,
%``Strange Matter,''
Phys.\ Rev.\ D {\bf 30}, 2379 (1984).
%%CITATION = PHRVA,D30,2379;%%

%\cite{Weber:2004kj}
\bibitem{Weber:2004kj}
F.~Weber,
%``Strange quark matter and compact stars,''
Prog.\ Part.\ Nucl.\ Phys.\  {\bf 54}, 193 (2005)
[arXiv:astro-ph/0407155].
%%CITATION = ASTRO-PH 0407155;%%

%\cite{Madsen:1998uh}
\bibitem{Madsen:1998uh}
J.~Madsen,
%``Physics and astrophysics of strange quark matter,''
Lect.\ Notes Phys.\ {\bf 516}, 162 (1999)
[arXiv:astro-ph/9809032].
%%CITATION = ASTRO-PH 9809032;%%

%\cite{Haensel:1986qb}
\bibitem{Haensel:1986qb}
P.~Haensel, J.~L.~Zdunik and R.~Schaeffer,
%``Strange Quark Stars,''
Astron.\ Astrophys.\  {\bf 160}, 121 (1986).
%%CITATION = AAEJA,160,121;%%

%\cite{Alcock:1986hz}
\bibitem{Alcock:1986hz}
C.~Alcock, E.~Farhi and A.~Olinto,
%``Strange Stars,''
Astrophys.\ J.\  {\bf 310}, 261 (1986).
%%CITATION = ASJOA,310,261;%%

%\cite{Glendenning:1997wn}
\bibitem{Glendenning:1997wn}
N.~K.~Glendenning,
``Compact stars: Nuclear physics, particle physics, and general relativity,''
(2nd edition), Springer, New York (2000).
%\href{http://www.slac.stanford.edu/spires/find/hep/www?irn=4025415}{SPIRES entry}

%\cite{Weber:1999qn}
\bibitem{Weber:1999qn}
F.~Weber,
``Pulsars as astrophysical laboratories for nuclear and particle physics,''
IOP Publishing (1999).
%\href{http://www.slac.stanford.edu/spires/find/hep/www?irn=4144937}{SPIRES entry}

%\cite{Madsen:1989pg}
\bibitem{Madsen:1989pg}
J.~Madsen,
%``Astrophysical Limits On The Flux Of Quark Nuggets,''
Phys.\ Rev.\ Lett.\  {\bf 61}, 2909 (1988).
%%CITATION = PRLTA,61,2909;%%

%\cite{Friedman:1990qz}
\bibitem{Friedman:1990qz}
J.~L.~Friedman and R.~R.~Caldwell,
%``Evidence Against A Strange Ground State For Baryons,''
Phys.\ Lett.\ B {\bf 264}, 143 (1991).
%%CITATION = PHLTA,B264,143;%%

%\cite{Aguilar:2002ad}
\bibitem{Aguilar:2002ad}
M.~Aguilar {\it et al.}  [AMS Collaboration],
%``The Alpha Magnetic Spectrometer (AMS) on the International Space
%Station. I:
%Results from the test flight on the space shuttle,''
Phys.\ Rept.\  {\bf 366}, 331 (2002)
[Erratum-ibid.\  {\bf 380}, 97 (2003)].
%%CITATION = PRPLC,366,331;%%

%\cite{AMS:2004}
\bibitem{AMS:2004}
C.H.~Chung {\it et al.}  [AMS Collaboration],
%``AMS on ISS. Construction of a particle physics detector on the
%International Space Station,''
Nucl.\ Inst.\ Meth.\ A\ submitted (2004).

%\cite{Sandweiss:2004bu}
\bibitem{Sandweiss:2004bu}
J.~Sandweiss,
%``Overview of strangelet searches and Alpha Magnetic Spectrometer: When will
%we stop searching?,''
J.\ Phys.\ G {\bf 30}, S51 (2004).
%%CITATION = JPHGB,G30,S51;%%

%\cite{Madsen:2001bw}
\bibitem{Madsen:2001bw}
J.~Madsen,
%``Color-flavor locked strangelets and their detection,''
J.\ Phys.\ G {\bf 28}, 1737 (2002)
[arXiv:hep-ph/0112153].
%%CITATION = HEP-PH 0112153;%%

%\cite{Madsen:2001fu}
\bibitem{Madsen:2001fu}
J.~Madsen,
%``Color-flavor locked strangelets,''
Phys.\ Rev.\ Lett.\  {\bf 87}, 172003 (2001)
[arXiv:hep-ph/0108036].
%%CITATION = HEP-PH 0108036;%%

%\cite{Rajagopal:2000ff}
\bibitem{Rajagopal:2000ff}
K.~Rajagopal and F.~Wilczek,
%``Enforced electrical neutrality of the color-flavor locked phase,''
Phys.\ Rev.\ Lett.\  {\bf 86}, 3492 (2001)
[arXiv:hep-ph/0012039].
%%CITATION = HEP-PH 0012039;%%

%\cite{Madsen:2000kb}
\bibitem{Madsen:2000kb}
J.~Madsen,
%``Intermediate mass strangelets are positively charged,''
Phys.\ Rev.\ Lett.\  {\bf 85}, 4687 (2000)
[arXiv:hep-ph/0008217].
%%CITATION = HEP-PH 0008217;%%

%\cite{Heiselberg:1993dc}
\bibitem{Heiselberg:1993dc}
H.~Heiselberg,
%``Screening in quark droplets,''
Phys.\ Rev.\ D {\bf 48}, 1418 (1993).
%%CITATION = PHRVA,D48,1418;%%

%\cite{Berger:1986ps}
\bibitem{Berger:1986ps}
M.~S.~Berger and R.~L.~Jaffe,
%``Radioactivity In Strange Quark Matter,''
Phys.\ Rev.\ C {\bf 35}, 213 (1987).
%%CITATION = PHRVA,C35,213;%%

%\cite{Madsen:2002iw}
\bibitem{Madsen:2002iw}
J.~Madsen and J.~M.~Larsen,
%``Strangelets as cosmic rays beyond the GZK-cutoff,''
Phys.\ Rev.\ Lett.\  {\bf 90}, 121102 (2003)
[arXiv:astro-ph/0211597].
%%CITATION = ASTRO-PH 0211597;%%

%\cite{Kalogera:2003tn}
\bibitem{Kalogera:2003tn}
V.~Kalogera {\it et al.},
%``The Cosmic Coalescence Rates for Double Neutron Star Binaries,''
Astrophys.\ J.\  {\bf 601}, L179 (2004)
[arXiv:astro-ph/0312101].
%%CITATION = ASTRO-PH 0312101;%%

%\cite{Lee:2002nk}
\bibitem{Lee:2002nk}
W.~H.~Lee, W.~Kluzniak and J.~Nix,
%``Binary coalescence of a strange star with a black hole: Newtonian
%results,''
Acta Astron.\  {\bf 51}, 331 (2001)
[arXiv:astro-ph/0201114].
%%CITATION = ASTRO-PH 0201114;%%

%\cite{Kluzniak:2002dm}
\bibitem{Kluzniak:2002dm}
W.~Kluzniak and W.~H.~Lee,
%``The swallowing of a quark star by a black hole,''
Mon.\ Not.\ Roy.\ Astron.\ Soc.\  {\bf 335}, L29 (2002)
[arXiv:astro-ph/0206511].
%%CITATION = ASTRO-PH 0206511;%%

%\cite{Prakash:2003em}
\bibitem{Prakash:2003em}
M.~Prakash and J.~M.~Lattimer,
%``A tale of two mergers: Searching for strangeness in compact stars,''
J.\ Phys.\ G {\bf 30}, S451 (2004)
[arXiv:astro-ph/0305306].
%%CITATION = ASTRO-PH 0305306;%%

%\cite{Gieseler}
\bibitem{Gieseler}
U.~D.~J.~Gieseler, T.~W.~Jones, and H.~Kang,
%``Time dependent cosmic-ray shock acceleration with self-consistent
%injection,''
Astron.\ Astrophys.\  {\bf 364}, 911 (2000).

%\cite{Longair:1994wu}
\bibitem{Longair:1994wu}
M.~S.~Longair,
``High-energy astrophysics. Vol. 2: Stars, the galaxy and the interstellar
medium,'' Cambridge University Press (1994).
%\href{http://www.slac.stanford.edu/spires/find/hep/www?irn=3524353}{SPIRES entry}

%\cite{Longair:1992ze}
\bibitem{Longair:1992ze}
M.~S.~Longair,
``High-energy astrophysics. Vol. 1: Particles, photons and their detection,''
Cambridge University Press (1992).
%\href{http://www.slac.stanford.edu/spires/find/hep/www?irn=2810697}{SPIRES entry}

%\cite{Schlickeiser}
\bibitem{Schlickeiser}
R.~Schlickeiser,
``Cosmic Ray Astrophysics'',
Springer-Verlag (Berlin-Heidelberg, 2002).

%\cite{Gleeson:1968}
\bibitem{Gleeson:1968}
L.~J.~Gleeson and W.~I.~Axford,
Astrophys.\ J.\  {\bf 154}, 1011 (1968).

%\cite{Benvenuto:1989hw}
\bibitem{Benvenuto:1989hw}
O.~G.~Benvenuto and J.~E.~Horvath,
%``Strangelet Seeding From Stellar Evolution: A Chain Without End?,''
Mod.\ Phys.\ Lett.\ A {\bf 4}, 1085 (1989).
%%CITATION = MPLAE,A4,1085;%%

%\cite{Medina-Tanco:1996je}
\bibitem{Medina-Tanco:1996je}
G.~A.~Medina-Tanco and J.~E.~Horvath,
%``The acceleration and propagation of strangelets in the galaxy: Numerical
%simulations,''
Astrophys.\ J.\  {\bf 464}, 354 (1996).
%%CITATION = ASJOA,464,354;%%

\end{thebibliography}
\end{document}